\documentclass[a4paper,12pt]{article}
\usepackage{epsfig}

\begin{document}

\begin{center}
{\Large{\bf $\sigma$ meson mass and width at finite density}}
\end{center}

\vspace{0.4cm}

\begin{center}
{\large{M.J. Vicente Vacas and E. Oset}}
\end{center}

\begin{center}
{\small{\it Departamento de F\'{\i}sica Te\'orica and IFIC, \\
Centro Mixto Universidad de Valencia-CSIC, \\
Ap. Correos 22085, E-46071 Valencia, Spain}}
\end{center}

\vspace{0.8cm}

\begin{abstract}
The $\sigma$ meson mass and width are studied at finite baryonic
density in the framework of  a chiral unitary approach which 
successfully reproduces the meson meson phase shifts and 
generates the $f_0$ and $\sigma$ resonances in vacuum.
\end{abstract}

\section{Introduction}

 There is mounting evidence showing the existence of a light scalar isoscalar 
meson  \cite{Groom:2000in,Tornqvist:2000jt,Aitala:2001xu,Colangelo:2001df}.
However, its nature as a genuine meson state or as a resonance dynamically 
generated in the $\pi\pi$ scattering is still under debate. A review of the 
current status of the discussion can be found in ref.  \cite{Ishida:2001pb}.

 In any case, it is such a broad resonance that its effects are hardly visible 
in any phase shifts or decay plots. The situation could be quite different at 
finite densities and/or temperatures where the $\sigma$ could become 
lighter and narrow \cite{Brown:1996qt,Hatsuda:2002ka}. Apart from its 
implications related to chiral symmetry restoration, this topic is of special 
interest due to the relevance of the $I=J=0$ channel in the nucleon-nucleon 
interaction. Any substantial change of this meson properties could alter the 
in medium nucleon-nucleon interaction and therefore our current understanding 
of the nuclear matter in heavy ion collision, neutron stars or even in normal 
nuclei if the changes occur at low densities.

The density dependence of the $\sigma$ properties has been studied in several models.
Bernard et al.  \cite{Bernard:1987im,Bernard:1988sx,Bernard:1988db}
have found a decrease of the mass as a function of the density using 
a generalized Nambu Jona-Lasinio model, until it converges to the mass of the
pion, its chiral partner. It is noticeable, that in the same model the
$\rho, \omega$ and $\pi$ mesons have almost constant masses.

In ref.  \cite{Hatsuda:1999kd}, Hatsuda et al. studied the $\sigma$ propagator
in the linear $\sigma$ model and found an enhanced and narrow spectral 
function  near the $2\pi$ threshold  caused by the partial restoration
of the chiral symmetry, where the $m_\sigma$ mass would approach $m_\pi$.
 The same conclusions were reached using the nonlinear chiral lagrangians 
in ref. \cite{Jido:2001bw}.

Similar results, with large enhancements in the $\pi\pi$ amplitude
 around the  $2\pi$ threshold,
 have been found in a quite different approach by studying the
$s-wave$, $I=0$ $\pi\pi$ correlations in nuclear matter
 \cite{Schuck:1988jn,Rapp:1996ir,Aouissat:1995sx,Chiang:1998di}. In these cases
the modifications of the $\sigma$ channel are induced by the strong $p-wave$
coupling of the pions to the particle-hole ($ph$) and $\Delta$-hole ($\Delta h$)
nuclear excitations. 
It was pointed out in  \cite{Aouissat:2000ss,Davesne:2000qj} that this
attractive $\sigma$ selfenergy induced by the $\pi$ renormalization in the
nuclear medium could be complementary to additional $s-wave$
renormalizations of the kind discussed in  \cite{Hatsuda:1999kd,Jido:2001bw}
calling for even larger effects.

On the experimental side, there are also several results showing strong medium
effects in the $\sigma$ channel at low invariant masses in the $A(\pi,2\pi)$ 
 \cite{Bonutti:1996ij,Bonutti:1998zw,Camerini:1993ac,bonutti,Starostin:2000cb}
and $A(\gamma,2\pi)$  \cite{metag} reactions. At the moment, the cleanest 
signal probably corresponds to the $A(\gamma,2\pi^0)$ reaction, which shows
large density effects that had been predicted in both shape and size in
ref. \cite{Roca:2002vd}, using a model for the $\pi\pi$ final state interaction
along the lines of the present work.

Our aim in this paper is to study the $\sigma$ mass and width at finite
densities in the context of the model developed in 
 \cite{Dobado:1990qm,Dobado:1993ha,Oller:1998ng,Oller:1999hw,Oller:1999zr,Oller:1997ti}.
These works, which provide a very economical and successful description of a wide
range of hadronic phenomenology, use as input the lowest orders lagrangian of chiral
perturbation theory  \cite{Gasser:1985ux} and calculate meson-meson scattering 
in a coupled channels unitary way. The $\sigma$ meson is not a basic
ingredient on this approach, it is dynamically generated and appears as a 
pole of the $\pi \pi$ scattering amplitude in the second Riemann sheet. 

The nuclear medium effects on the scalar isoscalar channel  were implemented
in this framework in refs.  \cite{Chiang:1998di,Oset:2000ev}. As in other
approaches, large medium effects were found. Namely, the imaginary part of the 
$\pi \pi$ scattering amplitude showed a clear shift of strength towards
low  energies as the density increases. 

In the next section we present, for the sake of completeness, a brief 
description of the model used for the $\pi \pi$ interaction in both vacuum 
and dense medium, which is already published elsewhere. The following section 
describes the method used to calculate the $\pi \pi$ scattering amplitude
in the complex energy plane 
and to search for the $\sigma$ pole position at finite densities.

\section{$\pi \pi$ interaction}

In this work we consider only the scalar isoscalar ($\sigma$) channel
and  follow the  simple method of ref.  \cite{Oller:1997ti}
for $\pi \pi$ interaction in vacuum and refs.  \cite{Chiang:1998di,Oset:2000ev}
for the nuclear medium effects. Additional information on this and related 
approaches for different spin isospin channels can be found in refs. 
 \cite{Oller:1997ti,Oller:1998ng,Oller:1999hw,Nieves:2000bx}.

\subsection{Vacuum}

 The basic idea is to solve a 
Bethe Salpeter (BS) equation, which guarantees unitarity, matching the low
energy results to the chiral perturbation theory ($\chi PT$) predictions.
We consider two coupled channels, $\pi \pi$ and $K \bar{K}$ and
neglect the $\eta \eta $ channel which is not relevant at the low energies 
we are interested in. 
 
\begin{figure}
%[htb]
 \epsfig {figure=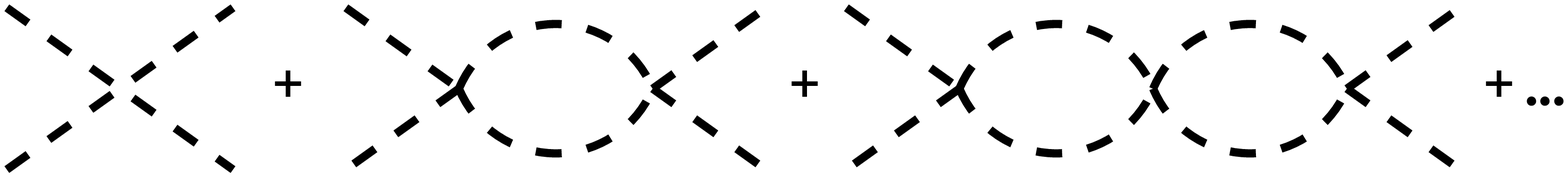,width=12.cm}
 \caption{Diagrammatic representation of the Bethe-Salpeter equation.}
 \label{fig:BSF}
\end{figure}

    The BS equation is given by 
\begin{equation}
\label{eq:BS}
T=V+VGT.
\end{equation}   
Eq. \ref{eq:BS}  is a matrix integral equation  which 
involves the two mesons one loop divergent integral, (see Fig.~\ref{fig:BSF}),  
where  $V$ and $T$ appear off shell. However, for this channel both functions 
can be factorized on shell out of the integral. The remaining off shell part 
can be absorbed by a renormalization of  the coupling constants
as it was shown in refs. \cite{Oller:1997ti,Nieves:1999hp}. 
 Thus, the BS equation becomes purely algebraic and the 
$VGT$  originally inside the loop integral becomes then the product 
of $V$, $G$ and $T$, with $V$ and $T$ the on shell amplitudes independent
of the integration variables, and  $G$  given by  the  expression  
\begin{equation}
G_{ii}(P) = i \int \frac{d^4 q}{(2 \pi)^4}
\frac{1}{q^2 - m_{1i}^2 + i \epsilon} \; \; 
\frac{1}{(P - q)^2 - m_{2i}^2 + i \epsilon}
\end{equation}
where $P$ is the momentum of the meson-meson system. This 
integral is regularized with a cut-off adjusted to optimize the fit
to the $\pi-\pi$ phase shifts ($\Lambda=1.03$ GeV).
The potential $V$ appearing in the BS equation is taken from 
the lowest order chiral Lagrangian
\begin{equation}
L_2 = \frac{1}{12 f^2} < (\partial_\mu \Phi \Phi - \Phi \partial_\mu
\Phi)^2 + M  \Phi^4 \, >
\end{equation}

\noindent
where the symbol $< >$ indicates the trace in flavour space,
$f$ is the pion decay constant and $\Phi$, $M$ are the pseudoscalar meson 
and mass $SU(3)$ matrices.
This model reproduces well phase shifts and inelasticities up to about 
1.2 GeV. The $\sigma$ and $f_0 (980)$ resonances appear as poles of the 
scattering amplitude. The coupling of channels is essential to produce the 
$f_0 (980)$ resonance, while the $\sigma$ pole is little
affected by the coupling of the pions to $K \bar{K}$
 \cite{Oller:1997ti}. 

\subsection{The nuclear medium.}

As we are mainly interested in the low energy region, which is not very
sensitive to the kaon channels, we will only consider the nuclear medium 
effects on the pions. The main changes of the pion propagation in the 
nuclear medium come from the p-wave selfenergy, produced basically 
by the the coupling of pions to particle-hole ($ph$) and Delta-hole 
($\Delta h$) excitations. For a pion of momentum $q$ it is given by 
\begin{equation}
\label{eq:self}
  \Pi(q)= {{\left({D+F}\over{2f}\right)^2 \vec q\,^2 U(q)}
            \over
           {1-\left({D+F}\over{2f}\right)^2 g' U(q)}}
\end{equation}
with $g'=0.7$ the Landau-Migdal parameter, $U(q)$ the  Lindhard function and
$(D+F)=1.257$. The expressions for the Lindhard functions  are taken
from ref. \cite{Oset:1990ey}. 

\begin{figure}[htb]
 \begin{center}
\epsfig{height=2.2cm,width=12.1cm,angle=0, figure=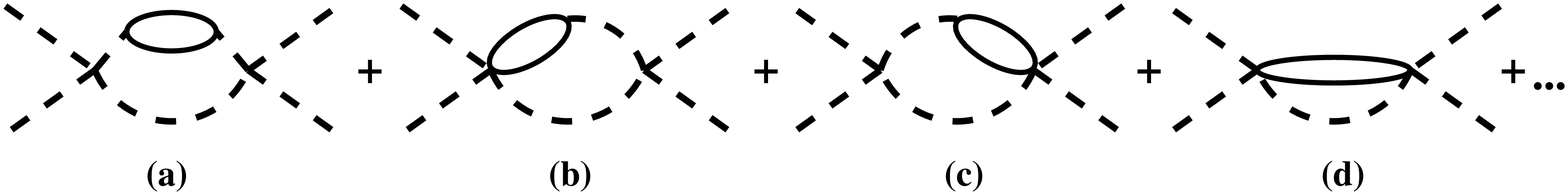}
\caption{Terms of the meson-meson scattering amplitude accounting for
$ph$ and $\Delta h$ excitation.}
 \label{fig:BSF2}
 \end{center}
\end{figure}

Thus, the in medium BS equation  will include the 
diagrams of Fig. \ref{fig:BSF2} where the solid line bubbles represent
the $ph$ and $\Delta h$ excitations.

In fact, as it was shown in  \cite{Chanfray:1999nn}, the contact terms with
the $ph$ ($\Delta h$) excitations of diagrams (b)(c)(d) cancel the 
off shell contribution  from the meson meson vertices in the term of Fig.
\ref{fig:BSF2}(a). 
Hence, we just need to calculate the diagrams of the free type 
(Fig. \ref{fig:BSF}) and those of Fig. \ref{fig:BSF2}(a) with the amplitudes
factorized on shell. Therefore, at first order in baryon density, we are left 
with simple meson propagator corrections which can be readily incorporated by 
changing the meson vacuum propagators by the in medium ones.

\begin{figure}[htb]
 \begin{center}
\epsfig{height=2.5cm,width=5.cm,angle=0,figure=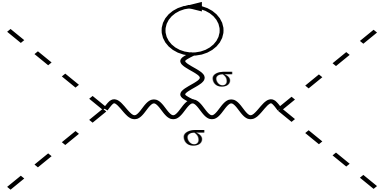}
 \caption{ Tadpole $\sigma$ selfenergy diagram.}
 \label{fig:SIG}
 \end{center}
\end{figure}

There are some other medium corrections, like the one 
 depicted in Fig. \ref{fig:SIG}
which has been considered in several papers like 
 \cite{Hatsuda:1999kd,Kunihiro:1999mt,Jido:2001bw} using the linear sigma model. 

\begin{figure}[htb]
 \begin{center}
\epsfig{height=2.0cm,width=12.2cm,angle=0,figure=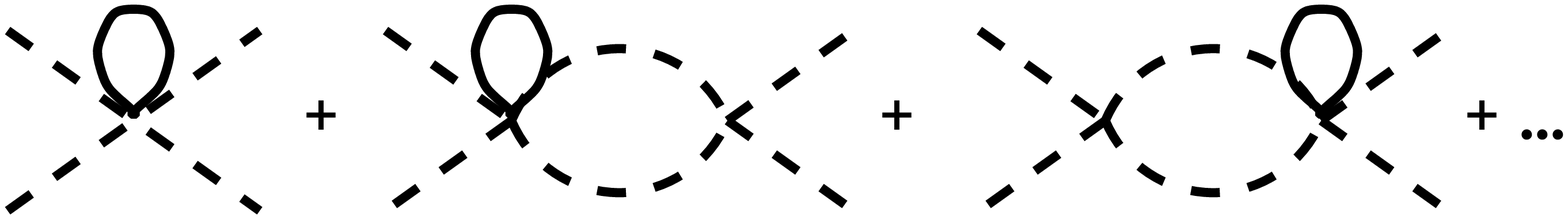}
 \caption{ Some tadpole diagrams contributing to the meson-meson scattering 
 amplitude }
 \label{fig:SIGPT}
 \end{center}
\end{figure}

In our approach, we start with only pseudoscalar meson fields. As it was
discussed previously, the sigma is 
generated dynamically  through the rescattering of the pions. A possible   
analog to the diagram of Fig. \ref{fig:SIG}  is given by the diagrams of 
Fig. \ref{fig:SIGPT}.

 We evaluate the contribution of these diagrams by using the
chiral Lagrangians
 \cite{Pich:1995bw,Meissner:1993ah,Bernard:1995dp,Ecker:1995gg} 
involving the octet of baryons and the octet of 
pseudoscalar mesons. We find  these  terms  to be  proportional to 
$\bar{p}\gamma^\mu p-\bar{n}\gamma^\mu n$ and thus they vanish in 
symmetric nuclear matter.

\begin{figure}[htb]
 \begin{center}
\epsfig{height=3.0cm,width=5.5cm,angle=0,figure=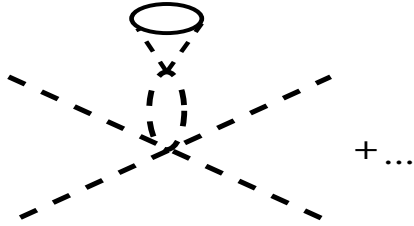}
 \caption{Higher order tadpole diagram. }
 \label{fig:SIGPT2}
 \end{center}
\end{figure}

Also we would need to consider diagrams of the type shown in Fig. \ref{fig:SIGPT2}
We get a negligible contribution to the the $\pi \pi$
scattering in the nuclear medium from this diagram  \cite{Oset:2000ev} and for
simplicity we do not include it in this work. 

\begin{figure}[htb]
 \begin{center}
\epsfig{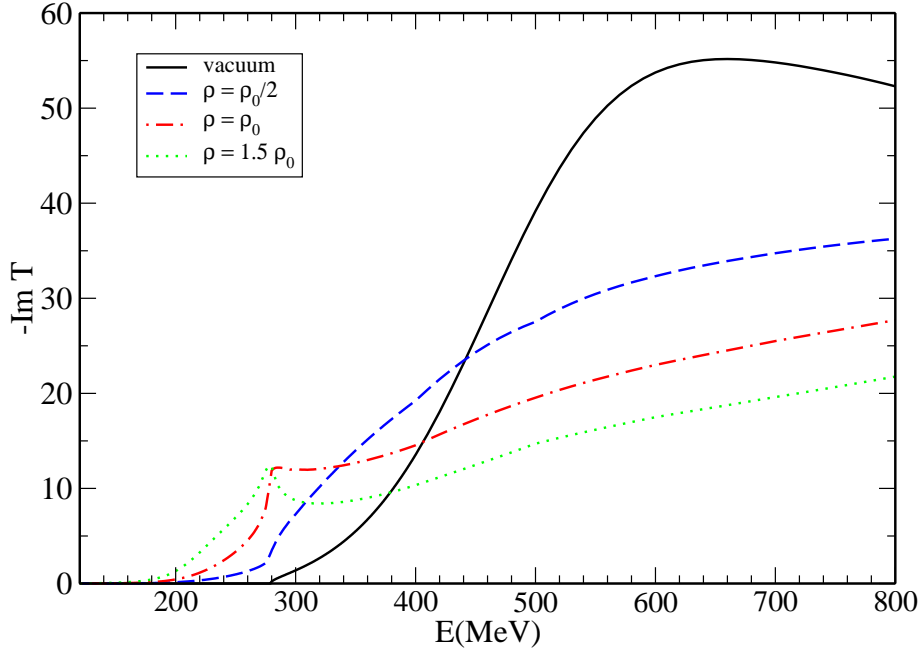}
 \caption{Imaginary part of the $\pi\pi$ scattering amplitude at several
 densities.}
 \label{fig:MED1}
\end{center}
\end{figure}

 The $\pi\pi$ scattering amplitude  obtained using this model
exhibits a strong shift towards low energies. In Fig. \ref{fig:MED1}, we show
the imaginary part of this amplitude for several densities. Quite similar
results have been found using different models 
 \cite{Aouissat:1995sx} and it has been suggested
that this accumulation of strength, close to the pion threshold, could reflect
a shift of the $\sigma$ pole which would approach the mass of the pion. 
Other pion selfenergy contributions related to $2ph$ excitations, and thus 
proportional to $\rho^2$, can be incorporated in the pion propagator. As we are
most interested in the region of low energies we can take as estimation the
corresponding piece of the optical potentials obtained from pionic atoms data,
following the procedure of ref. \cite{Chiang:1998di} and substituting in Eq.
\ref{eq:self}
\begin{equation}
\left({D+F}\over{2f}\right)^2  U(q)
\;\longrightarrow\; 
\left({D+F}\over{2f}\right)^2  U(q) -4\pi C_0^* \rho ^2
\label{eq:self2}
\end{equation}
with $\rho$ the nuclear density and $C_0^*=(0.105+i 0.096) m_\pi^{-6}$.
Its effects are small except at large densities as can be appreciated by
comparing  Fig. \ref{fig:MED1}, with Fig. 7 of ref. \cite{Chiang:1998di}
where this piece is included.
\section{$\sigma$ meson mass and width as a function of the baryon density} 

In this section we will consider the lightest pole position of the
$\pi\pi$ scattering amplitude in the S=I=0 channel in the complex energy 
plane, which corresponds to the $\sigma$ resonance. In our model, the vacuum 
pole position occurs at a mass around 500 MeV and the width is around 400 MeV. 
 A compilation of values for these magnitudes obtained in several modern 
analyses can be found in ref.  \cite{Markushin:2001kx}. See also Fig. 1 of 
ref. \cite{Tornqvist:2000jt} and \cite{vanBeveren:2002vw}. 

The analytical structure of the meson meson scattering amplitude is driven 
by the meson loops, which are given by the formula
\begin{equation}
  G_{\pi \pi } = i\int ^{\infty }_{0}
  {{d^{4}q}\over{(2\pi)^4}} 
  \, D_{\pi }(q^{0},\vec{q}\,)D_{\pi }
 (\sqrt{s}-q^{0},-\vec{q}\,) 
\end{equation}
where $D_{\pi }$ is the $\pi$ propagator which, in the nuclei, incorporates 
the p-wave coupling of the pions to particle-hole and $\Delta$-hole 
excitations.
We can simplify the calculation by using the Lehmann representation for the 
meson propagators
\begin{equation}
 D_{\pi }(q^{0},\vec{q}\,)=-\frac{2}{\pi }\int ^{\infty }_{0}
dx\, \frac{x\, \, Im\, D_{\pi }(x,\vec{q}\,)}
{(q^{0})^{2}-x^{2}+i\epsilon }
\end{equation}
After some algebraic manipulation we obtain
\begin{equation}
\label{eq:poles}
 G_{\pi \pi }=\int ^{\infty }_{0}\frac{dE}{2\pi }
(\frac{1}{P^{0}-E+i\epsilon }-\frac{1}{P^{0}+E-i\epsilon })\, F(E) 
\end{equation}
where $F(E)$ is a real function given by
\begin{equation}
 F(E)=\int {{d^{3}q}\over{(2\pi)^3}}\, 
 \int ^{E}_{-E}{{dx}\over{\pi}}\: Im\, D_{\pi }((E+x)/2,
\vec{q}\,)\: Im\, D_{\pi }((E-x)/2,\vec{q}\,) 
\end{equation}
which in vacuum takes the simple form
\begin{equation}
F_{vac}(E)=\frac{p_{\pi }}{4\pi \, E} 
=\frac{1}{4\pi} \sqrt{\frac{1}{4}-\frac{m_{\pi }^{2}}{E^2}} 
\end{equation}
As we are interested in the positive energy region,
we can concentrate our attention on the first term
of Eq. \ref{eq:poles} which is the only one that has a cut. 
We can rewrite that term, neglecting the $\epsilon$, like
\begin{equation}
\frac{F(E)}{P^{0}-E}\;
\rightarrow\;
\frac{F(Real(P^0))}{P^{0}-E}+
\frac{F(E)-F(Real(P^0))}{P^{0}-E}
\end{equation}
The $E$ integration of the first term it is the only piece that depends 
on the Riemann sheet and can be readily evaluated. In the second sheet
the integral takes the value
\begin{equation}
 \int ^{\infty }_{0,(II)} dE 
(\frac{F(Real(P^0))}{P^{0}-E})=
\int ^{\infty }_{0,(I)} dE 
(\frac{F(Real(P^0))}{P^{0}-E})+2 \pi i F(Real(P^0))
\end{equation}
Substituting this result in Eq. \ref{eq:poles} we get the following expression
for the two meson propagator in the second sheet
\begin{equation}
G_{\pi \pi }^{(II)}=i F(Real(P^0))+\int ^{\infty }_{0}\frac{dE}{2\pi }
(\frac{1}{P^{0}-E}-\frac{1}{P^{0}+E})\, F(E) 
\end{equation}
The rest of the pieces entering the BS equation are analytical single valued 
functions and therefore we can proceed to look for poles in the scattering
amplitude.

\section{Results}
\begin{figure}[htb]
 \begin{center}
\psfig{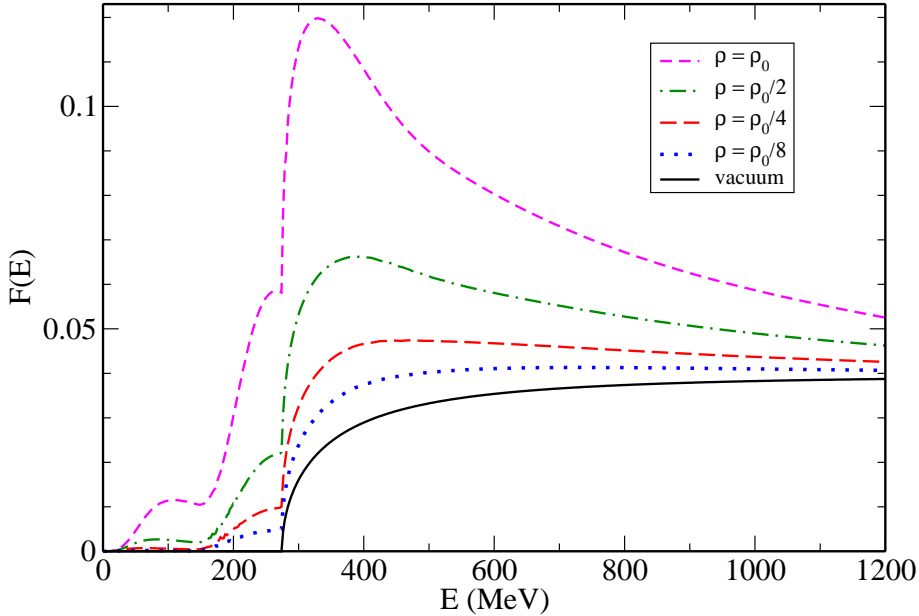}
\caption{$F(E)$ at several densities.}
  \label{fig:R8}
 \end{center}
\end{figure}

The function $F(E)$ includes the phenomenological information on the
pion selfenergy in the nuclear medium. The results for several densities are
shown in Fig. \ref{fig:R8}.

\begin{figure}[htb]
 \begin{center}
\psfig{height=3.cm,figure=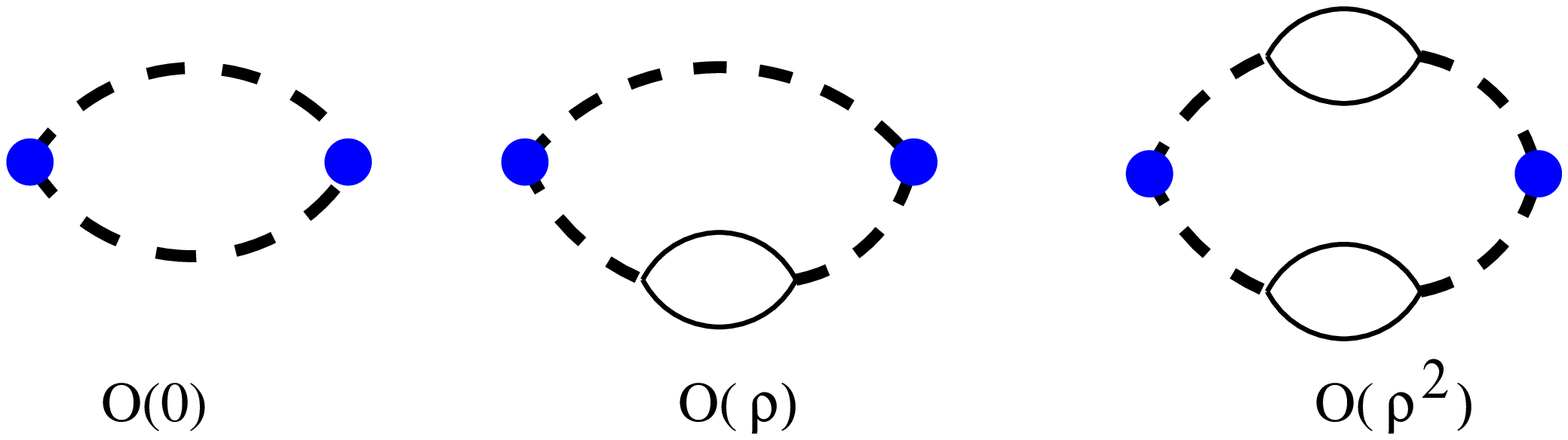}
\caption{$\sigma$ decay channels classified according to their density
dependence}
  \label{fig:R9}
 \end{center}
\end{figure}

The $F(E)$ strength is related to the imaginary part of the $\sigma$ 
propagator and therefore reflects  the energy and density dependence of the
different $\sigma$ decay channels. We can start classifying them according 
to their density dependence (See. Fig.\ref{fig:R9}).

At low densities, the $\sigma$ meson can only decay into two pions,
 $\sigma \rightarrow \pi+\pi$. Therefore we have a threshold at $E=2m_\pi$.
As the density grows the  $\sigma \rightarrow \pi+(Nh)$ decay channel 
becomes relevant and we have a new threshold at $E=m_\pi$, which is clearly
visible in the curve corresponding to  $\rho=\rho_0/8$. Finally, at larger
densities, mechanisms such as $\sigma \rightarrow 2(Nh)$
which go like $\rho^2$  become important. They are possible even at very low
energies.

\begin{figure}[htb]
 \begin{center}
\psfig{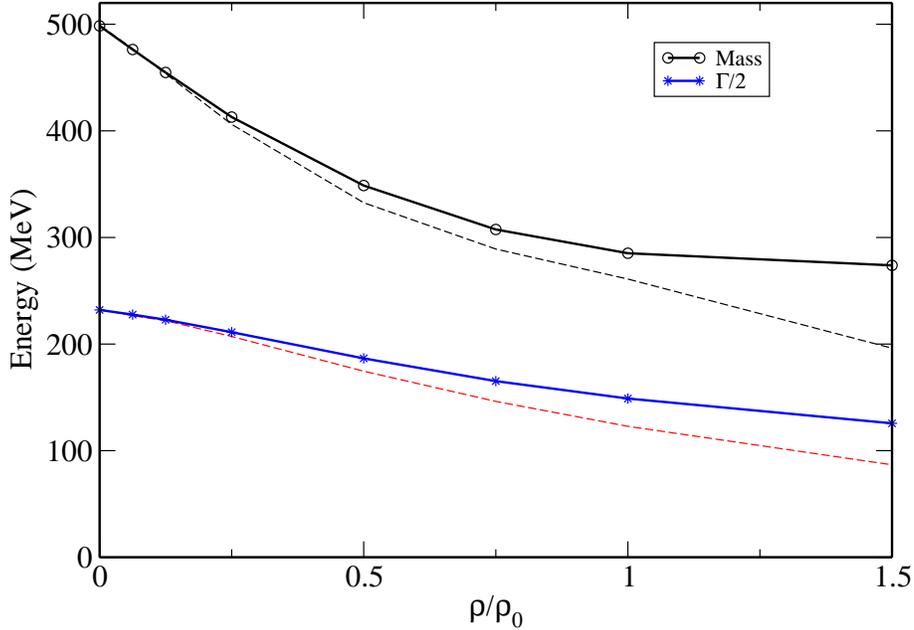}
\caption{$\sigma$ mass  and half width as a function
of the density. Dashed lines include also the $2ph$ pion selfenergy pieces}
  \label{fig:R10}
 \end{center}
\end{figure}

Using the two meson propagator of Eq. \ref{eq:poles} we can solve the
BS equation, obtain the meson-meson scattering amplitude
 and look for the poles in the complex energy plane. The 
results for the $\sigma$ pole position are shown in Fig. \ref{fig:R10} 
for densities up to 1.5 $\rho_0$. Note, however, that the calculation is more 
reliable at low densities because some contributions of order $\rho^2$ or 
higher are missing. 
 We find that both mass and width decrease as the density increases,
reaching a mass around 250 MeV and a similar width at 1.5 times the nuclear
density. At large densities the $2ph$ pieces of the pion selfenergy become
relevant decreasing further both mass and width.
These results are little sensitive to the kaon channel. Its omission affects 
less than 1 percent the mass and increases the width around 5 percent.
The results of Fig. \ref{fig:R10} can be cast in terms of an effective potential
which can be approximated by
\begin{equation}
V_\rho=a \,\frac{\rho}{\rho_0} +b\, \left(\frac{\rho}{\rho_0} \right)^2
\end{equation}
with $a=-358-i\, 108$ MeV and $b=140+i\, 23.6$ MeV.

Qualitatively similar results for the mass are found in other models. 
See for instance Fig. 10 of ref.  \cite{Hatsuda:2001da}. However,  we 
should stress that there the changes respond to a reduction of the pion decay 
constant ($f$) value which we have kept constant. The basic ingredient that 
drives the mass decrease in our calculation is the p-wave interaction of the 
pion with the baryons in the medium.

  Although the $\sigma$ mass drops significantly, the width stays relatively
large, even when the mass is close or below the $2\pi$ threshold.  This width 
comes mostly from medium decay channels, namely,
the decay into a single pion and a $ph$ excitation or $2ph$ excitations. 
 Therefore we should not expect any signal of narrow $\sigma$ mesons
in the medium as expected in \cite{Hatsuda:2002ka} or in  models that include
only purely mesonic decays.
 Nonetheless the low mass could modify the long and medium range
nucleon nucleon interaction in nuclei or nuclear matter.

\section{Conclusions}
In this work, we have searched for the lightest pole of the $\pi\pi$ scattering 
amplitude, the $\sigma$ meson,  in the complex energy plane as a function of 
the baryon density. 

The growing consensus that the $\sigma$  is rather a $\pi\pi$ resonance 
of dynamical origin than a genuine QCD state built from $q\bar q$ pairs,
finds its support in recent chiral unitary formulations of the $\pi\pi$
interaction which we have followed in the present work.
This approach has been extended to include the interaction of the pions with a
nuclear medium, which allows us to trace the density dependence of the 
$\sigma$ pole.

A quite general agreement has been reached about these medium modifications of
the  $\pi\pi$ interaction between different groups, either working in the 
present approach or
using other $\pi\pi$  scattering formulations, as long as they satisfy some 
minimal chiral constraints. All these models are, however, involved numerically
to the point that extrapolating analytically the results  to the complex
plane is a difficult task, particularly when some pieces of the $\pi$
selfenergy, like the $2p2h$ contributions, are taken from a comparison with
pionic atoms data and not from a theoretical model.

In this work, the analytical extrapolation was made possible by the use of the
Lehmann representation for the  $\pi$ propagators. This allowed us to express the
scattering matrix in terms of the pion propagators evaluated on the real energy
axis.

The results obtained show a dropping of the $\sigma$ mass as a function of the
density, down to values close to two pion masses at normal nuclear density. The
width decreases moderately and reaches values  around  300 MeV at $\rho=\rho_0$.
Yet, this decrease  is surprising in view of the fact that in the nuclear medium
there are more channels for the $\sigma$ decay, like pion-$ph$ or 2$ph$. 
The presence of these channels partly makes up for the strong reduction of the 
$2\pi$ channel due to the smaller phase space available when the $\sigma$ 
mass drops.

Altogether, the changes in the mass and width of the $\sigma$ are quite large, 
compared to other typical meson or baryon properties in a dense medium.
Therefore, we expect drastic signals in the invariant mass distributions of
two pions (or photons) in 2$\pi$ (2$\gamma$) production experiments if those
pions (photons) can couple strongly to the scalar isoscalar channel. 
Some present data seem to support these claims, although further experiments and
calculations should be performed to test  these findings.

\section{Acknowledgements}
This work is partly supported by DGICYT contract no. 
BFM2000-1326.

\end{document}